\documentclass{procDDs}
\usepackage{color}

\title{Semi-classical quantization rules for a periodic orbit of hyperbolic type}

\author{Hanen LOUATI}
{Universit\'e de Tunis El-Manar, D\'epartement de Math\'ematiques, 1091 Tunis, Tunisia.
Aix Marseille Univ, Univ Toulon, CNRS, CPT, Marseille, France}                 
{louatihanen42@yahoo.fr}                                   
\coauthor{\bf Michel ROULEUX}                       
{Aix Marseille Univ, Univ Toulon, CNRS, CPT, Marseille, France}               
{rouleux@univ-tln.fr}                                               

\def\const{\mathop{\rm const.}\nolimits}
\def\e{\mathop{\rm \varepsilon}\nolimits}
\def\id{\mathop{\rm Id}\nolimits}
\def\im{\mathop{\rm Im}\nolimits}

\def\re{\mathop{\rm Re}\nolimits}
\def\Sp{\mathop{\rm Sp}\nolimits}
\def\supp{\mathop{\rm supp}\nolimits}
\begin {document}
\newtheorem{theo}{Theorem}
\maketitle

\index{Author1, I.I.}                              
\index{Author2, I.I.}                              
\index{Coauthor, I.I.}                             %

\begin{abstract}
Determination of periodic orbits for a Hamiltonian system together with their semi-classical quantization
has been a long standing problem.
We consider here resonances for a $h$-Pseudo-Differential Operator $H(y,hD_y;h)$
induced by a periodic orbit of hyperbolic type at energy $E_0$. 
We generalize the framework of \cite{GeSj}, in the sense that
we allow for both hyperbolic and elliptic eigenvalues of Poincar\'e map, and show that 
all resonances in $W=[E_0-\e _0,E_0+\e _0]-i]0,h^\delta]$, $0<\delta<1$, are given by a 
generalized Bohr-Sommerfeld quantization rule (BS). 
\end{abstract}

\section{Hypotheses and the main result.}

Though our results apply to some  manifolds, we restrict the attention to ${\bf R}^n$. 
Let $H(y,hD_y;h)$ be a self-adjoint $h$-Pseudo-Differential Operator ($h$-PDO) on $L^2({\bf R}^n)$
\begin{equation}
\label{1.1}
\begin{aligned}
H^w&(y,hD_y; h)u(y;h)=(2\pi h)^{-n}\int\int e^{i(y-y')\eta'/h}\cr
&\times H(\frac{y+y'}{2},\eta';h)u(y')\, dy'\, d\eta'\cr
\end{aligned}
\end{equation}
We assume it has Weyl symbol $H(y,\eta;h)\in S^0(m)$, where $m$ is an order function (for example $m(y,\eta)= (1+|\eta|^2)^M$), and
\begin{equation*}
\begin{aligned}
S^N&(m)=\{H\in C^\infty(T^*{\bf R}^n): \forall\alpha\in{\bf N}^{2n}, \exists C_\alpha>0,\cr
&|\partial^\alpha_{(y,\eta)}H(y,\eta;h)|\leq C_\alpha h^N m(y,\eta)\}\cr
\end{aligned}
\end{equation*}
with the semi-classical expansion
$H(y,\eta; h) \sim H_0(y,\eta)+hH_1(y,\eta)+\cdots, h\rightarrow 0$.
Here $H_0$ is the principal symbol of $H$, $H_1$ its sub-principal symbol. We assume that
$H(y,\eta;h)+i$ is elliptic, i.e. $|H(y,\eta;h)+i|\geq \const m(y,\eta)$, and extends as an analytic symbol in the sense of \cite{Sj1} in a sector
$$\Gamma_0=\{(y,\eta) \in T^*{\bf C}^n : |\im (y,\eta)|\leq\const \langle\re (y,\eta)\rangle\}$$
Let the energy surface $H_0^{-1}(E_0)$ be regular for some $E_0\in{\bf R}$, that we may set up to 0. So the Hamiltonian vector field
$X_{H_0}$ has no fixed point on $H_0^{-1}(0)$, hence on nearby energy surfaces $H_0^{-1}(E)$. 
Let $\Phi^t = \exp(tX_{H_0}) : T^\ast {\bf R}^n \rightarrow T^\ast {\bf R}^n$ and
\begin{equation}
\label{1.2}
K_E =\{\rho\in H_0^{-1}(E), \Phi^{t}(\rho)\nrightarrow\infty \ \hbox{as} \ |t|\to\infty\}
\end{equation}
be the trapped set at energy $E$. We assume that 
$K_0=\gamma_0$ is a periodic orbit of period $T_0$, or possibly a finite union of such orbits.
Let ${\cal P}_0$ be Poincar\'e map, or the first return map, acting on Poincar\'e sections 
$\Sigma(\rho)\subset T^*{\bf R}^n$, $\rho\in\gamma_0$.
Assume also that $\gamma_0$ is non degenerate, i.e. 1 is not an eigenvalue of $d{\cal P}_0$, then it follows from
Poincar\'e Continuation Theorem that for $|E|$ small enough,
$\gamma_0$ belongs to a one parameter family (the center manifold)
of such periodic orbits $K_E=\gamma_E$ of period $T_E$. We identify 
the center manifold with a neighborhood of the zero-section in $T^*{\bf S}^1$, and 
each $\Sigma(\rho)$, $\rho\in\gamma_0$ with
$\Sigma\approx T^*{\bf R}^d$, modulo the action of Hamiltonian flow. 
Both $\overline{\gamma}$ and $\Sigma$ are symplectic manifolds. For $\rho\in\gamma_0$, let
$\lambda_j, 1\leq j\leq 2d=2(n-1)$ be the eigenvalues of $A_0=d{\cal P}_0:{\bf C}^{2d}\rightarrow{\bf C}^{2d}$
(Floquet multipliers). 

The space ${\bf C}^{2d}$ has the orthogonal symplectic decomposition in (generalized) eigenspaces $F_\lambda$
relative to the family $(\lambda_j)_{1\leq j\leq 2d}$. We are interested in the case where 
$A_0$ is partially hyperbolic, i.e. has at least one eigenvalue $\lambda$ of modulus $\ne 1$.
Assume also that Poincar\'e map is non degenerate, i.e. $F_1 = \{0\}$, and also $F_{\lambda} = \{0\}$ for all $\lambda \leq 0$.
We say that $\lambda\in {\bf C}$ is {\it elliptic} (ee for short) if
$|\lambda|=1$ ($\lambda \neq \pm 1$) and {\it hyperbolic} (he) if $|\lambda| \ne 1$; 
if moreover $\lambda\in{\bf R}$ we call it {\it real hyperbolic} (hr)
and {\it complex-hyperbolic} (hc) otherwise.
Under the last assumption we can define $B_0=\log A_0$. Eigenvalues 
$\mu=\mu(\lambda)=\log\lambda$ of $B_0$ (Floquet exponents) verify $\mu(\overline{\lambda})=\overline{\mu(\lambda)}$.
Accordingly, exponent $\mu$ is said ee if $\re\mu=0$, hr if $\mu\in{\bf R}\setminus0$, and
hc if $\mu\in{\bf C}\setminus{\bf R}$. So eigenvalues of $B_0$ have the form
$\mu_j,-\mu_j$ in the hr sector, $\mu_j,\overline\mu_j$ in the ee sector,  
and $\mu_j, -\mu_j,\overline{\mu_j},-\overline{\mu_j}\neq 0$, $\re \mu_j>0$,
(with same multiplicity) in the hc sector. Every pair of simple Floquet exponents $\mu_j,-\mu_j$ in the ee sector
can be ordered thanks to the symplectic structure on ${\bf C}^{2d}$. Namely if $u$ is an eigenvector corresponding to such a $\mu$,
then ${1\over2i}\sigma(\overline u,u)\in{\bf R}\setminus0$ with a sign independent of $u$; so we call $\mu$ an {\it eigenvalue of the 
first kind} if ${1\over2i}\sigma(\overline u,u)>0$. Alternatively, the Hamiltonian flow 
of $b(\rho) = {1\over2}\sigma(\rho, B_0\rho)$ (Hermitian form) with time of negative
imaginary part is ``expanding'' on 
an eigenvector corresponding to an eigenvalue of the first kind.
Let $r$ be the number of distinct $\mu_j$'s. 
For simplicity, assume $r=d$, hence $b$ diagonalizable. We know \cite{Br1} that
in a suitable basis $b(\rho)$ is a linear combinaison of elementary quadratic polynomials $Q_j$, and that 
we can choose complex symplectic coordinates, so that $Q_j$ takes the form $Q_j=x_j\xi_j$.
The $Q_j$'s play an important r\^ole, since they are formally  ``transverse eigenvectors'' for $H$, microlocalized near $\gamma_0$.

Our next Hypothesis is relative to {\it partial hyperbolicity} of Poincar\'e map,
in the sense that there exists $j\in\{ 1,...,r\}$, such that $\re\mu_j>0$.
For hyperbolic dynamical systems, we know \cite{A} that there are generically elliptic elements. 
Let $F_{\mu_{j}}$, $\re \mu_j\geq0$ denote again the eigenspace associated with $\mu_j$. 
We can rewrite the decomposition of
${\bf C}^{2d}$ in the sum of unstable $F^+=\bigoplus_{j=1} ^{r=d} F_{\mu_{j}}$, resp. stable $F^-=\bigoplus_{j=1} ^{r=d} F_{-\mu_{j}}$ spaces
where $F^\pm \simeq {\bf C}^d$ are (complex) Lagrangian subspaces of ${\bf C}^{2d}$, and $F^+$ contains the eigenspaces associated with
eigenvalues of the first kind.

Our last Hypothesis concerns the non-resonance condition relative to Floquet exponents, which is required to achieve
Birkhoff normal form, namely
\begin{equation}
\label{1.7}
\forall k\in {\bf Z}^d: \sum_{j=1}^{d} k_j\mu_j\in 2i\pi{\bf Z}  \Longrightarrow  
\sum_{j=1}^{d} k_j \mu_j = 0
\end{equation}
For instance, when $n=2$ and $\mu_1 = i\omega_1$, it means that the rotation number $\omega_1$ is irrational.
We need also the strong non-resonance condition on Floquet exponents: 
\begin{equation}
\label{1.8}
\forall k\in{\bf Z}^d: \sum_{j=1}^{d} k_j \mu_j \in 2i\pi {\bf Z} \Longrightarrow k=0
\end{equation}
\noindent {\it Examples}: 1) The Model Hamiltonian 
\begin{equation}
\label{1.9}
\widetilde H(hD_t,x,hD_x;h)=-hD_t+\sum_{j=1}^d\mu_jQ_j^w(x,hD_x)
\end{equation}
$Q_j^w(x,hD_x)={1\over2}(x_jhD_{x_j}+hD_{x_j}x_j)$, with Periodic Boundary Conditions on ${\bf S}^1\times{\bf R}^d$,
serves as a guide-line throughout this work. Here $x$ may denote complex variables, in some Bargmann representation
of the Hamiltonian.

2) A physical example is given by
$H(y,hD_y)=-h^2\Delta_y+|y|^{-1}+ay_1$ on ${\bf R}^n$ (repulsive Coulomb potential perturbed by Stark effect) near an energy level
$E>2/\sqrt a$. More generally, Schr\"odinger operators with potentials with two or more bumps, 
which are semi-classical analogues of Helmholtz operators with obstacles, are considered in \cite{Sj3}. Such a periodic orbit is sometimes
called a {\it libration} in the case of a compact energy surface.

3) The geodesic flow on the one-sheeted hyperboloid in ${\bf R}^3$ has an (unstable) periodic orbit of hyperbolic type 
(Poincar\'e example). This example generalizes (\cite{Chr}, App.C) to a surface of revolution in ${\bf R}^4$, involving two
symmetric periodic orbits of real-hyperbolic type at energy $E_1$, and 
a periodic orbit of mixed type in between, at energy $E_2<E_1$ (with real-hyperbolic and elliptic elements). 
Our method carries to the case when $H(y,hD_y)$ 
is the geodesic flow on such manifolds, and $K_E$ the union of two periodic orbits, provided we modify Grushin operator of \cite{SjZw}
accordingly.

We are concerned with semi-classical {\it resonances} of $H$ near 0, in the framework of ``complex scaling''
theory and its extensions \cite{ReSi}, \cite{HeSj}, i.e. the discrete spectrum in the lower-half plane of some suitable analytic continuation of $H$
as a closed, Fredholm, but non-selfadjoint operator.
Our main result, the generalized Bohr-Sommerfeld quantization condition, can be formulated as follows (see also \cite{LoRo}):

\begin{theo}
Under the hypotheses above, let (after re-ordering) $\mu_j=i\omega_j$, $j=1,\cdots,\ell$,
$\omega_j>0$ be the set of elliptic Floquet exponents for $H_0$. Recall $H_1$ from (\ref{1.1}), and let $H_1(y(t),\eta(t))\,dt$ 
the sub-principal 1-form. We define the {\rm semi-classical} action along $\gamma_E$, by
${\cal S}(E;h)=S_0(E)+hS_1(E)+{\cal O}(h^2)$ with
\begin{equation}
\label{1.10}
S_0(E)=\frac{1}{2\pi}\int_{\gamma_E}\xi\,dx
\end{equation}
\begin{equation}
\label{1.11}
\begin{aligned}
S_1&(E)=-\frac{1}{2\pi }\int_0^{T(E)} H_1(y(t),\eta(t))\,dt+\cr
&\frac{1}{4i\pi}\sum_{j=1}^d\mu_j(E)+{g_\ell\over4}\cr
\end{aligned}
\end{equation}
Here $\mu_j(E)=\mu_j+{\cal O}(E)$ is Floquet exponent at energy $E$, 
$g_\ell\in{\bf Z}$ Gelfand-Lidskii (or Cohnley-Zehnder) index of $\gamma_E$ (depending
only on elliptic elements, see below). Then given any $0<\delta<1$,
the resonances of $H$ near 0 in the ``window'' $W_h=[-\e _0,\e _0]-i]0,h^\delta]$ ($\e _0>0$ sufficiently small)
are given (at first order in $h$) by the generalized BS quantization condition 
\begin{equation}
\label{1.12}
{\cal S}(E;h)+{1\over2i\pi}\bigl(\sum_{j=1}^d k_j\mu_j(E)+{\cal O}(h|k|^2)\bigr)=mh
\end{equation}
with $m\in{\bf Z}, k\in{\bf Z}^d$, 
provided $|m|h\leq \e_0$, $|k|h\leq \const h^\delta$. 
\end{theo}

In the elliptic case, a similar theorem (for real spectrum) was obtained in \cite{Ba}, \cite{BaLa}, and \cite{Ra};
in the real hyperbolic case, in \cite{GeSj}, \cite{GeSj1} for $|\im E|={\cal O}(h)$, and \cite{Sj2} in dimension 2 with $|\im E|={\cal O}(h^\delta)$
or even larger, but selecting a single Floquet parameter in the semi-classical Floquet decomposition
of $H$ near $\gamma_0$,
i.e. few ``longitudinal'' or ``principal'' quantum numbers $m\in{\bf Z}$.
For related results about trace formulas or concentration
of eigenvalues in the compact case, see \cite{Vo1}, \cite{Vo2}, \cite{SjZw},
\cite{NoSjZw}, \cite{Chr}. For the wave equation outside convex obstacles, where all Floquet exponents of the 
billiard map are real-hyperbolic, see
\cite{Ik}, \cite{Ge}. 

It is well known \cite{SjZw}, \cite{NoSjZw}, \cite{FaLoRo} that all resonances in $W_h$ are given by the zeroes of a ``$\zeta$-function''
$\zeta(z;h)=\det(\id-N(z;h))$, where 
$N(z;h)=\Pi_N(h)M(z;h)^*\Pi_N(h)+{\cal O}(h^N)$ is the approximation of order $N$ of a semi-classical {\it monodromy operator}  $M(z;h)^*$
in a suitable (diagonal) basis of homogeneous polynomials. We compute here the semi-classical action at first order, which is the
main contribution of $\zeta(z;h)$. We make an essential use of Birkhoff normal form in the neighborhood of $\gamma_0$.
Higher approximations could be obtained as in the 1-D case \cite{IfRo}. 

\section{Construction of the monodromy operator}

The main object to be constructed is the monodromy operator $M^*(E;h)$, a $h$-FIO defined on a Poincar\'e section and
quantizing Floquet operator
associated with the periodic orbit.\\ 

\noindent {\it 1) Birkhoff normal form (BNF)}.\\

We start to find suitable coordinates near $\overline\gamma$. When $\re\mu_j>0$ for all $1\leq j\leq d$, the stable/unstable
manifold theorem guarantees the existence of involutive manifolds $\Gamma_\pm$ in a neighborhood of $\gamma_0$ with
$T_\rho\Gamma_++T_\rho\Gamma_-+T_\rho\overline\gamma=T^*{\bf R}^n$, $\rho\in\gamma_0$. 
There are (real) symplectic coordinates
$(t,\tau,x,\xi)$ such that
$\xi=0$, $d\xi\neq0$ on $\Gamma_+$, $x=0$, $dx\neq0$ on $\Gamma_-$, and $(t,\tau)$ parametrize $\overline\gamma$.
Intersecting with the energy surfaces $H_0^{-1}(E)$ gives the splitting 
$T_\rho\Gamma_+(E)+T_\rho\Gamma_-(E)+T_\rho\gamma(E)=T_\rho^*H_0^{-1}(E)$, $\rho\in\gamma(E)$,  
and $\Gamma_\pm(E)$ are Lagrangian
submanifolds in $T^*{\bf R}^n$ intersecting transversally along $\gamma(E)$. In these coordinates 
$H_0(y,\eta)=f(\tau)+\langle B(t,\tau,x,\xi)x,\xi\rangle$.
Here $f$ parametrizes the energy parameter
$f(\tau)=E$, and is related with the period $T(E)$ of $\gamma(E)$ by 
$f'(\tau)={2\pi\over T\circ f(\tau)}$. Performing a first canonical transformation gives $B(t,\tau,x,\xi)=\widetilde B_0+{\cal O}(|\tau|,|x,\xi|)$,
where the eigenvalues of $\widetilde B_0$ are Floquet exponents for $d{\cal P}_0$ with positive real part. 
When $\mu_k$ is an eigenvalue of the first kind, splitting of $T_\rho^*H_0^{-1}(E)$
as above still holds provided we take complex variables. 
In both cases however, under the non resonance conditions (\ref{1.7}), (\ref{1.8}) BNF holds in the classical sense \cite{Br2} as well as in the
semi-classical sense \cite{GuPa} and takes, modulo a small remainder term,
operator $H^w(y,hD_y;h)$ to a polynomial in $hD_t$ and $Q_j^w(x,hD_x)$,
$Q_j(x,\xi)$ as above. In particular, the principal part of $H^w(y,hD_y;h)$ is
in BNF is given by (\ref{1.9}) in some suitable Bargmann (still formal) representation, provided a reparametrization of energy.\\

\noindent {\it 2) Microlocalisation in the complex domain}.\\

Resonances here are considered from the point of vue of analytic dilations
and Lagrangian deformations; taking into account that there exists an {\it escape function} (that grows along the flow of $X_{H_0}$)
which implies kind of a ``virial condition'' outside the trapped set
$\gamma_0$, the 
most relevant region of phase-space for such deformations is a neighborhood of $\gamma_0$.
Here we make a complex scaling of the form $(x,\xi)\mapsto(e^{i\theta}x,e^{-i\theta}\xi)$, followed also by a small
deformation in the $(t,\tau)$ variables. Our main tool is
FBI transformation (metaplectic FIO with complex phase, see \cite{Sj1}, \cite{M}) which takes the form, in coordinates 
$(s,y;t,x)\in T^*{\bf R}^n\times T^*{\bf C}^n$  adapted to $\Gamma_\pm$ as in BNF, and acts on $u\in L^2({\bf R}^n)$ as
$$T_0u(x,h)=\int e^{i\varphi_0(t,s;x,y)/h}u(s,y)\,ds\,dy$$ 
where $\varphi_0(t,s;x,y)=\varphi_1(t,s)+\varphi_2(x,y)$,
$\varphi_1(t,s)={i\over2}(t-s)^2$, $\varphi_2(x,y)={i\over2}\bigl[(x-y)^2-{1\over2}x^2\bigr]$.
The corresponding canonical transformation is $\kappa_0=(\kappa_1,\kappa_2)$, with
$\kappa_1:(s,-\partial_s\varphi_1)\mapsto(t,\partial_t\varphi_1)$, $\kappa_2:(y,-\partial_y\varphi_2)\mapsto(x,\partial_x\varphi_2)$,
and the corresponding pluri-subharmonic (pl.s.h.) weight $\Phi_0=\Phi_1+\Phi_2$, with 
$\Phi_1(t)=(\im t)^2/2$, $\Phi_2(x)=|x|^2/4$.
In a very small neighborhood of $\gamma_0$, whose size
will eventually depend on $h$, corresponding to $\theta=-\pi/4$, and that we call the ``phase of inflation'', it turns out
that $H^w(y,hD_y;h)$ takes the simple form above whose principal term is given in (\ref{1.9}), and the corresponding weight 
$\widetilde\Phi(t,x)$ is just
$\Phi_0(t,x)$. Otherwise we take $\theta$ small enough in a somewhat larger neighborhood of $\gamma_0$, which we call
the ``linear phase''. Farther away from $\gamma_0$ (in the ``geometric phase'') the weight is implied by the 
escape function given by the non-trapping condition. All these weights are
patched together in overlapping regions,
so to define a globally pl.s.h. function in ${\bf C}^n$. They also
define the contour integral for writing a $h$-FIO in the complex domain, using
almost analytic extensions because of the loss of analyticity in BNF. After these transormations, $H_0-E$ becomes elliptic
everywhere but on $\gamma_E$.\\

\noindent {\it 3) Poisson operator and its normalisation}.\\
 
Let ${\bf R}_t^n$ be the section $\{t\}\times{\bf R}^d$ of ${\bf R}^n$ (in BNF coordinates). 
We look for $K(t,E):L^2({\bf R}^d)\to L^2({\bf R}_t^n)$ (formally), microlocalized near $\Gamma_+(E)$, 
such that $H(hD_t,x,hD_x;h)K(t,E)=0, \ K(0,E)=\id$. 
In the ``phase of inflation'', considering realizations in the complex domain adapted to the weight $\widetilde\Phi_\theta$,
it takes the form $K(t,E)v(x;h)=\int\int e^{i(S(t,x,\eta)-y\eta)/h}a(t,x,\eta;E,h)v(y)\,dy\wedge d\eta$. 
Solving eikonal and transport equations, we find that the leading term of $S$ and $a$ with respect to BNF is given by
those of (\ref{1.9}), and $K(t,E)$ is also in BNF. Let
$\chi\in C^\infty({\bf R})$, be equal to 0 near 0, 1 near $[2\pi,\infty[$.
There is a $h$-PDO $B(E)=B^w(x,hD_x;E)$ such that 
$L(t,E)=K(t,E)B(E)$ satisfies the ``flux norm'' identity of \cite{SjZw}
\begin{equation}
\label{2.3}
\bigl({i\over h}[H,\chi(t)]L(t,E)v|L(t,E)v\bigr)=\bigl(v|v\bigr)
\end{equation}
Considering also microlocal reproducing kernels,
we show there exists $P(s,E)=\id_{L^2({\bf R}^n_s)}+{\cal O}(h)$ such that
\begin{equation}
\label{2.4}
\begin{aligned}
&\int dE\,L(\cdot,E)P(\cdot,E)L^*(E)=\id_{L^2({\bf R}^n)}\cr 
&\int dE\,L^*(E)L(s,E)P(s,E)=\id_{L^2({\bf R}^n_s)}\cr
\end{aligned}
\end{equation}
These (formal) computations can be carried out in the framework of FIO's in the complex domain.\\

\noindent {\it 4) The monodromy operator}.\\

We set $K_0(t,E)=K(t,E)$ where $K(t,E)$ is Poisson operator with Cauchy data at $t=0$, and $L_0(t,E)=K_0(t,E)B(E)$; we set similarly 
$L_{2\pi}(t,E)=K_0(t-2\pi,E)B(E)$ with Cauchy data at $t=2\pi$. The monodromy operator (or semi-classical Poincar\'e map) is defined by 
\begin{equation}
\label{2.5}
M^*(E)=L_{2\pi}^*(E){i\over h}[H,\chi]L_0(\cdot,E)
\end{equation} 
as an operator on $L^2({\bf R}^d)$.
As a function de $\chi$, $M^*(E)$ follows a ``0-1 law'': it is 0 if $\supp \chi\subset]0,2\pi[$, and unitary if 
$\chi$ equals 0 near 0, and 1 near $2\pi$. Unitarity is shown using (\ref{2.3}) and (\ref{2.4}). For the model case one has 
$M^*(E)v(x)=e^{-2i\pi E/h}e^{\pi \mu}v(xe^{2\pi \mu})$ when $\int\chi'(t)\, dt=1$.
Moreover Schwartz kernel of $M^*(E)$ given by
\begin{equation}
\label{2.6}
{\cal M}^*(t,x,z)=\chi'(t)\int e^{i\Psi_2(t,x,z,\eta')/h}m^*\,d\eta'
\end{equation}
(where $m^*=m^*(t,x,z,\eta';h)$ is a suitable symbol) is in BNF, and the integral is independent of $t$. In fact
$M^*(E)=e^{iR^w(x,hD_x;E,h)/h}$,
where $R$ is $h$-PDO in BNF, self-adjoint for real $E$. 

\section{The quantization condition}

Reducing the spectral problem for $H^w(x,hD_x;h)$ through a Grushin operator as in \cite{SjZw}, \cite{NoSjZw}, \cite{FaLoRo}, 
we consider the approximate kernel of $M^*(E)-\id_{L^2({\bf R}^d)}$, after
taking suitable analytic extensions with respect to $E$ of Poisson and monodromy operators. 
The discrete set of $E$'s such that 1 belongs to the spectrum of $M^*(E)$ is precisely the set of resonances.
We follow \cite{GeSj}, \cite{Ra}, while taking further advantage of BNF to compute resonances lying farther from the real axis.
Since $M^*(E)$ is in BNF (formal) eigenfunctions of $M^*(E)$ are given by
homogeneous polynomials $f_k(x)$, whose degree $|k|$ depends on the accuracy of BNF; we can take $|k|=h^{-\delta}$,
for some $0<\delta<1$. So we need to compute $M^*(E)f_k$ by asymptotic stationary phase using (\ref{2.6}), the leading term being 
$f_k(x)$ times the phase factor $e^{i{\cal S}(E;h)/h}$ involving the semi-classical action ${\cal S}(E;h)=S_0(E)+hS_1(E)+{\cal O}(h^2)$, 
and times a half-density. Still we shall formally discuss the various terms in (\ref{1.10})-(\ref{1.12}), without resorting to $M^*(E)$. \\ 

\noindent {\it 1) Action integral and half-density}.\\

In (\ref{1.12}) we can readily identify the classical action $S_0(E)={1\over2\pi}\int_{\gamma_E}\xi\,dx$, as well as 
the principal part of a half-density.
These are the ``constant terms'' (independent of the 
transverse variables) in $M^*(E)f_k$ (the kernel of $M^*(E)$ being given in (\ref{2.6})), which
contribute to the ``longitudinal quantum number'' $m$ as in the 1-D case. Namely, looking for a WKB solution, we solve
$$(H^w-E)\bigl(a^E_\pm(y;E,h)\exp[i\phi^E_\pm(y,E)/h]\bigr)=0$$ 
and only take in account the ``outgoing'' solution (+). 
The solution to the eikonal equation $H_0(y,{\partial\phi^E\over\partial y})=E$, is a multivalued function
$\phi^E$, such that if we denote by ${\cal J}_E$ the 
``first return'' map (in space variable), then $\phi^E({\cal J}_Ey)=\phi^E(y)+C(E)$, where $C(E)=2\pi S_0(E)$ 
verifies $S'_0(E)=T(E)$. Solving next for the first transport equation, we find accordingly 
$a_0^E({\cal J}_Ey)=\rho(E)a_0^E(y)$, where $\rho(E)$ can be computed in the coordinates of Sect.4.1. 
Assume for simplicity that the subprincipal symbol $H_1(y,\eta)=0$. If 
$d{\cal P}_E=\begin{pmatrix}A(E)&0\\ 0&{}^tA(E)^{-1}\end{pmatrix}$ denotes linearized Poincar\'e map, we have 
$|\rho(E)|=|\det A(E)|^{-1/2}=|\prod_{j=1}^d\lambda_j(E)|^{-1/2}$, where $\lambda_j(E)$ belong to the set of eigenvalues of $d{\cal P}_E$
of the first kind (of modulus 1), or of modulus $>1$, see \cite{GeSj}, Prop.1.1. 
Thus in ${\cal S}(E;h)$ and (\ref{1.12}) we have identified $S_0(E)$ and the term $\frac{1}{4i\pi}\sum_{j=1}^d\mu_j(E)$ of $S_1(E)$.
We can also identify the contribution of polynomials $f_\alpha$ in $M^*(E)f_\alpha$.
Namely, let $D^E$ be the restriction of $d{\cal P}_E$ to $F_+$, and $D^E_*:{\cal Q}^N(F_+)\rightarrow{\cal Q}^N(F_+)$
its push-forward acting on the space of polynomials of degree $\leq N$. The eigenvalues of $D^E_*$ are then the numbers
$\lambda(E)^{-k}=\lambda_1(E)^{-k_1}\cdots\lambda_d(E)^{-k_d}$, with $|k|=k_1+\cdots+k_d$. 
This gives the ``transverse'' quantum numbers $k_j$ and the sum on the LHS of (\ref{1.12}), the remainder ${\cal O}(|k|^2)$
being given by higher order terms in BNF.\\

\noindent {\it 2) Cohnley-Zehnder index}.\\

We are left with the RHS of (\ref{1.12}).
We recall \cite{GL}, \cite{Ra}, \cite{CoZe}, \cite{SaZe} the index of a symplectic arc (Gelfand-Lidskii, or Conley-Zehnder), which 
appears in the quantization condition when elliptic elements occur. It
is defined for a differentiable path $\Psi:[0,T]\to\Sp(2n;{\bf R})$ such that 
$\Psi(0)=\id$ and $\det(\id-\Psi(T'))\neq0$ 
for some $T'\in[0,T]$. In the present case, let
$Z(s)$ solve the variational system along $\gamma_E$, i.e. $\dot Z(s)=JH''(\Phi^s(\rho))Z(s)$, $Z(0)=\id$, 
where we recall $\Phi^s(\rho)$ is the flow of $X_{H_0}$ issued from $\rho\in\gamma_E$ as in (\ref{1.2}). We define $\Psi(s)=d{\cal P}_E(s)$ 
as the co-restriction $\widetilde Z(s)$ of $Z(s)$ to Poincar\'e sections, i.e. $\Psi:[0,T]\to\Sp(2d;{\bf R})$. 
Then, Conley-Zehnder index can be interpreted as the mean winding number of the eigenvalues of the first kind,
or equivalently the number of such eigenvalues crossing 1. Here we take advantage of BNF and the splitting of $T^*{\bf R}^d$ 
in the incoming/outgoing manifolds to simplify the argument of \cite{SaZe}. 

Recall $\Lambda\in{\cal L}(2d)$ (the Grassmanian of Lagrangian planes), can be represented as
$\Lambda={\cal U}({\bf R}^d)$, ${\cal U}\in U(d)$ (unitary group); given 
a differentiable path ${\bf S}^1\to{\cal L}(2d), s\mapsto\Lambda_s={\cal U}_s({\bf R}^d)$, Maslov universal class is defined by
\begin{equation}
\label{3.1}
dm(\Lambda_s)=dm_s=\frac{1}{4\pi}d(\arg \det {\cal U}_s^2)
\end{equation}
Consider for simplicity the linear situation, i.e. Hamiltonian (\ref{1.9}), without loss of generality we may assume $T=T'=1$.
Integrating the variational system we find 
$\widetilde Z(s)=\begin{pmatrix}C(s)&B(s)\\ B(-s)&C(-s)\end{pmatrix}$, where 
${C(s)\choose B(s)}$, $C(s),B(s)$ symmetric, is a basis of the Lagrangian subspace
$\Lambda_+(s)=\exp sX_H(\Gamma_+(0))$, and ${B(-s)\choose C(-s)}$ (by Fourier transform)
a basis of $\Lambda_-(s)=\exp sX_H(\Gamma_-(0))$. 
As in \cite{DoRo} we introduce Cayley transformation
$${\cal U}(s)=(C(s)+iB(s))(C(s)-iB(s))^{-1}$$ 
which parametrizes $\Lambda_+(s)$ 
and ${\cal U}(s)$ is unitary except for finitely many points $s\in[0,1]$, corresponding to discontinuities of $dm_s$
at the passage of a caustics (i.e. whenever $s\omega_k\in2\pi{\bf Z}$, for some $s,k$, $\mu_k=i\omega_k$ being a Floquet
exponent of the first kind). 
Let $d\widetilde m_s$ be the regular part of $dm_s$, by a direct computation
$\frac{1}{4\pi}\int_0^{2\pi}d\widetilde m(\Lambda(s))=2\sum_{k=1}^\ell\omega_k$,
where the summation runs over elliptic elements. The variation of $dm_s$ at a caustics can be determined as in (\cite{Ho}, Sect.3.3),
using also the non degeneracy of Poincar\'e map. This gives (\ref{1.11}) in the model case. The general case follows from BNF and stability
of index $g_\ell$ under perturbation. Note that computing $M^*(E)f_k$ by stationary phase readily gives $g_\ell$ mod 4.

\end {document}